\documentclass[preprint,3p]{elsarticle}
\journal{MSSP}

\usepackage{graphicx}
\usepackage{amsmath}
\usepackage{amssymb}
\usepackage{xspace}
\usepackage[dvipsnames]{xcolor}
\usepackage{subcaption}
\usepackage{longtable,booktabs}

\usepackage{blindtext}
\usepackage{comment}
\usepackage{hyperref}

\newcommand{\nc}{\newcommand}
\nc{\rnc}{\renewcommand}
\nc{\bs}{\boldsymbol}
\rnc{\matrix}[2]{\left[\!\!\begin{array}{#1}
			#2\end{array}\!\!\right]}
\rnc{\vector}[1]{\matrix{c}{#1}}
\nc{\mm}[1]{\boldsymbol{#1}}
\nc{\mms}[1]{\boldsymbol{#1}}
\nc{\real}[1]{\Re\left\{ #1 \right\}}
\nc{\imag}[1]{\Im\left\{ #1 \right\}}
\nc{\dd}{\mathrm{d}}
\nc{\ii}{\mathrm{i}}
\nc{\ee}{\mathrm{e}}
\nc{\inv}{^{-1}} 
\nc{\herm}{^{\mathrm H}}
\nc{\tra}{^{\mathrm T}}
\nc{\conj}[1]{ \overline{#1} }
\nc{\diag}{\operatorname{diag}}
\nc{\CoR}{e}
\nc{\naft}{N}
\nc{\rlinh}{\hat{\mm r}}
\nc{\td}[1]{\breve{#1}}
\nc{\gtd}{\td{\mm g}}
\nc{\qqbtd}{\td{\qq}_{\mathrm b}}
\nc{\epsAL}{\varepsilon_{\mathrm{AL}}}
\nc{\epsDL}{\varepsilon_{\mathrm{DL}}}
\nc{\qq}{\mm q}
\nc{\uu}{\mm u}
\nc{\qqb}{\qq_{\mathrm b}}
\nc{\qqi}{\qq_{\mathrm i}}
\nc{\ui}{{\uu}_{\mathrm{i}}}
\nc{\dui}{\dot{\uu}_{\mathrm{i}}}
\nc{\ub}{{\uu}_{\mathrm{b}}}
\nc{\fex}{\mm f}
\nc{\fexb}{{\mm f}_{\mathrm b}}
\nc{\fexi}{{\mm f}_{\mathrm i}}
\nc{\Wb}{\mm W_{\mathrm b}}
\nc{\Wbtra}{\mm W\tra_{\mathrm b}}
\nc{\half}{\frac12}
\nc{\qqbk}{\qqb^k}
\nc{\qqik}{\qqi^k}
\nc{\qqikm}{\qqi^{k-1}}
\nc{\qqikp}{\qqi^{k+1}}
\nc{\qqbkm}{\qqb^{k-1}}
\nc{\qqbkp}{\qqb^{k+1}}
\nc{\ubk}{\ub^k}
\nc{\ubkm}{\ub^{k-1}}
\nc{\uikph}{\ui^{k+\half}}
\nc{\uikmh}{\ui^{k-\half}}
\nc{\fexh}{\hat{\mm f}}
\nc{\fexbh}{\hat{\mm f}_{\mathrm b}}
\nc{\fexih}{\hat{\mm f}_{\mathrm i}}
\nc{\qtil}{\tilde{\qq}}
\nc{\Mtil}{\tilde{\mm M}}
\nc{\Ktil}{\tilde{\mm K}}
\nc{\Dtil}{\tilde{\mm D}}
\nc{\eye}{\mm I}
\nc{\nmod}{N_{\mathrm{mod}}}
\nc{\ncon}{C}
\nc{\MOD}[1]{#1}
\nc{\ie}{i.\,e.\xspace}
\nc{\eg}{e.\,g.\xspace}
\nc{\cf}{cf.\,}
\nc{\myquote}[1]{`#1'}
\nc{\etal}{et al.\xspace}
\nc{\etc}{etc.\xspace}
\nc{\fabstand}{\,}
\nc{\fp}{\fabstand.}
\nc{\fk}{\fabstand,}
\nc{\tab}[5][tbh]{\begin{table}[#1]\centering\caption{#4\label{tab:#5}}\begin{tabular}{#2}\hline #3 \\ \hline\end{tabular}\end{table}}
\nc{\e}[2]{\begin{equation} #1 \end{equation}}
\nc{\est}[1]{\begin{equation*} #1 \end{equation*}}
\nc{\ea}[1]{
	\begin{eqnarray}
		#1\end{eqnarray}}
\nc{\east}[1]{
	\begin{eqnarray*}
		#1 \end{eqnarray*}}
\nc{\fref}[1]{{Fig.~\ref{fig:#1}}}
\nc{\frefo}[1]{{\ref{fig:#1}}}
\nc{\frefs}[1]{{Figs.~\ref{fig:#1}}}
\nc{\tref}[1]{{Tab.~\ref{tab:#1}}}
\nc{\trefo}[1]{{\ref{tab:#1}}}
\nc{\trefs}[1]{{Tab.~\ref{tab:#1}}}
\nc{\eref}[1]{{Eq.~(\ref{eq:#1})}}
\nc{\erefo}[1]{(\ref{eq:#1})}
\nc{\erefs}[1]{{Eqs.~(\ref{eq:#1})}}
\nc{\sref}[1]{{Section~\ref{sec:#1}}}
\nc{\srefo}[1]{\ref{sec:#1}}
\nc{\srefs}[1]{{Sections~\ref{sec:#1}}}
\nc{\ssref}[1]{{Subsection~\ref{sec:#1}}}
\nc{\ssrefo}[1]{\ref{sec:#1}}
\nc{\ssrefs}[1]{{Subsections~\ref{sec:#1}}}
\nc{\aref}[1]{{{Appendix~\ref{asec:#1}}}}
\nc{\arefo}[1]{{\ref{asec:#1}}}
\nc{\arefs}[1]{{{Appendices~\ref{asec:#1}}}}
\nc{\MATLAB}{{\sf{MATLAB}}\xspace}
\nc{\CalculiX}{{\sf{CalculiX}}}
\nc{\Julia}{{\sf{Julia}}\xspace}

\nc{\fts}[1]{\footnotesize{#1}}
\nc{\bts}[1]{\begin{itemize} #1 \end{itemize}}
\nc{\numbered}[1]{\begin{itemize} #1 \end{itemize}}
\nc{\note}[1]{\textcolor{BurntOrange}{#1}}
\nc{\blue}[1]{\textcolor{blue}{#1}}
\nc{\red}[1]{\textcolor{red}{#1}}
\nc{\gray}[1]{\textcolor{Gray}{#1}}
\nc{\myfig}[4]{
	\begin{figure}[#1]
		\centering
		\includegraphics[width=#4\textwidth]{#2}
		\caption{#3}\label{fig:#2}
	\end{figure}}

\nc{\mytab}[5][tbh]{\begin{table}[#1]\centering\caption{#4\label{tab:#5}}\begin{tabular}{#2}\hline #3 \\ \hline\end{tabular}\end{table}}

\nc{\rarrow}{\blue{$\boldsymbol{\rightarrow}\;$}}

\begin{document}

\begin{frontmatter}
    \title{Prediction and validation of the strongly modulated forced response of two beams undergoing frictional impacts}
    \author{Monjaraz Tec, C.$^1$}
    \author{Kohlmann, L.$^2$}
    \author{Schwarz, S.$^2$}
    \author{Hartung, A.$^2$}
    \author{Gross, J.$^1$}
    \author{Krack, M.$^1$}
    \address{$^1$ University of Stuttgart, GERMANY}
    \address{$^2$ MTU Aero Engines AG, GERMANY}

\begin{abstract}
We consider two cantilevered beams undergoing frictional impacts at the free end.
The beams are designed to be of similar geometry so that they have distinct but close natural frequencies.
Under harmonic base excitation near the primary resonance with the higher-frequency fundamental bending mode, the system shows a strongly modulated non-periodic response.
The purpose of this work is to \MOD{analyze to what extent the non-periodic vibro-impact dynamics can be predicted.}
\MOD{To this end, we use a recently developed modeling and simulation approach.}
The approach relies on component mode synthesis, the massless boundary concept and an appropriate time stepping scheme.
Unilateral contact and dry friction are modeled as set-valued laws and imposed locally within the spatially resolved contact area.
A linear model updating is carried out based on the natural frequencies and damping ratios identified in the regime without impacts.
The nonlinear simulation of the steady-state response to forward and backward stepped sine excitation is compared against measurements.
The results are in very good agreement, especially in the light of the uncertainty associated with the observed material loss in the contact region and the nonlinear behavior of the clamping.
\end{abstract}


\end{frontmatter}

\section{Introduction\label{sec:introduction}}
A vibro-impact process denotes any physical process where the mutual interaction between temporary (opening-closing) contacts and vibrations plays a central role \cite{Babitsky.2013}.
Technical examples are
machining processes (\eg pneumatic hammers, ultrasonic drillers),
impact vibration absorbers,
systems with backlash/freeplay (\eg hammering gears in drive trains, piping systems, turbine blades with hammering tip shrouds),
rotor-stator interactions (\eg between blades and casing of aero engines or within rotor bearings).

The prediction of vibro-impact processes is challenging:
To properly resolve the collision events, time steps are needed which are orders of magnitude smaller than the periods of the low-frequency vibration modes (typically of primary concern).
To properly resolve the highly concentrated stresses within the contact region, \MOD{a spatial discretization is needed whose length scale} is orders of magnitude smaller than the wavelengths of the critical vibration modes.
Each collision may initiate an elastic wave which propagates within the solid body, and, depending on the boundary conditions, may travel back and thus affect the contact interactions \cite{Seifried.2010}.
In multi-body dynamics, it is common practice to describe the loss of kinetic energy of rigid bodies by elastic wave effects (and possibly also plastic deformation) via a coefficient of restitution.
In the case of recurrent impacts between vibrating (flexible) bodies, such a decoupling of low- and high-frequency dynamics is generally not appropriate.
Instead, contact mechanics and elastodynamics need to be simultaneously resolved on the time scale of the vibration, making the prediction of vibro-impact processes computationally demanding \cite{wrig2006,john1989,Ibrahim.2009}.

Several works address the numerical prediction and experimental validation of vibro-impact processes \cite{Moon.1983,vandeVorst.1996,vandeVorst.1998,Wagg.2002,Krishna.2012,Elmegard.2014,Gandhi.2017}.
In most cases, slender bodies (usually beams) are considered and the impacts occur due to motion constraints, \ie, mechanical stops.
Good agreement between prediction and experimental results was achieved near primary resonances under harmonic excitation in the following works.
Wagg and Bishop \cite{Wagg.2002} considered a cantilevered beam with a (single-sided) rigid stop.
They modeled the system using modal truncation and Newton's impact law with an empirically determined coefficient of restitution.
Krishna and Padmanabhan \cite{Krishna.2012} and Elmegard \etal \cite{Elmegard.2014} considered a cantilevered beam with symmetric (double-sided) stops.
The system was modeled using finite elements with subsequent component mode synthesis in \cite{Krishna.2012} and using a single-degree-of-freedom approximation in \cite{Elmegard.2014}.
In both cases, the contact with the mechanical stop was modeled using a unilateral linear spring.
Gandhi \etal  \cite{Gandhi.2017} considered a cantilevered tapered beam with an end mass and (single-sided) elastic stop.
They modeled the system using modal truncation and a unilateral linear spring.
Remarkable agreement was achieved not only near primary but also sub- and super-harmonic resonances by van de Vorst \etal \cite{vandeVorst.1996,vandeVorst.1998}.
They considered a pinned-pinned beam in \cite{vandeVorst.1996} and a cantilevered beam in \cite{vandeVorst.1998}, in both cases subjected to a (single-sided) mechanical stop.
They modeled the system using finite elements with subsequent component mode synthesis, in combination with a unilateral Hertzian spring.
They showed that multiple degrees of freedom are needed to accurately describe even the low-frequency dynamics.
It was shown that higher modes may destabilize sub-harmonic resonances and cause additional super-harmonic resonances.

It is important to emphasize that the \emph{quantitative validation} in all the above mentioned studies is limited to \emph{periodic} vibration regimes.
Indeed, many studies do not even report non-periodic limit states.
This is almost surprising given that it is well-known that the strong impact nonlinearity often gives rise to non-periodic limit states, also near primary resonances.
For instance, this is experimentally demonstrated already in the seminal work of Moon and Shaw from the early 1980s \cite{Moon.1983}.
Besides the \MOD{coexisting periodic responses} also addressed in the aforementioned references, they encountered a non-periodic (chaotic) response, not far from the primary resonance.
Unfortunately, they only achieved qualitative agreement with their single-mode model \cite{Moon.1983}.

The main purpose of the present work is to analyze to what extent non-periodic vibrations of vibro-impact systems can be predicted.
To properly assess the predictive capabilities, we limit the model updating to the regime without impacts, and we do not calibrate empirical parameters (such as a coefficient of restitution) to the measured vibration response.
Besides the focus on non-periodic vibration regimes, there are further original aspects that distinguish the present validation from those reported in the literature:
First, we consider \MOD{both unilateral and frictional contact}, while the aforementioned studies designed the test rig to avoid frictional effects.
Second, we consider a nominally conforming flat-on-flat contact, as opposed to the Hertzian contacts typically considered in the literature.
Finally, we consider impacts between flexible bodies with close natural frequency, so that the dynamic deformation of more than one body is relevant for the system dynamics.

In \sref{testrig}, we present the test rig.
Modeling and simulation are described in \sref{model}, including model updating based on measurements in the regime without impacts.
Predictions and measurements of the (nonlinear) vibro-impact response are confronted in \sref{NLFR}.
This article ends with concluding remarks in \sref{conclusions}.

\section{Test rig\label{sec:testrig}}
\myfig{htbp}{testrig}
{Schematic of the test \MOD{rig}: overview with equipped instrumentation
}{0.99}
\myfig{htb}{rig_mesh}
{Model of test rig: (a) Schematic illustration of contact interface between upper probe with protrusion and lower probe \MOD{(front view)}, (b) contact mesh, (c) nodes involved in elastic layer defined for each bolted joint interface}{0.99}
%
A schematic illustration of the test rig including its instrumentation is given in \fref{testrig}.
The structure under test consists of two cantilevered beams clamped onto the armature/platform of a large electro-dynamic exciter (acting in the vertical direction in \fref{testrig}-left).
Near the tip of the beams, cylindrical material probes are attached to either beam.
In the absence of vibrations, there is a small clearance between the probes.
\MOD{
As stated in the introduction, many previous works considered Hertzian contacts.
These can be simply modeled using a single Hertzian spring.
In contrast, we consider a nominally flat-on-flat contact.
Such contacts are quite common in engineering.
They are more challenging to model.
In particular, the contact area has to be spatially resolved.
}
The considered contact area is relatively small, and its orientation ensures that friction is relevant (\fref{rig_mesh}a). 
More specifically, contact occurs between the protrusion of the upper probe with the flat surface of the lower probe (\fref{rig_mesh}a).
The surface of the protrusion is rectangular, with an aspect ratio of 7:10, and the longer edge amounts to only $2\%$ of the beams' length.
The lower beam has a uniform rectangular cross section, while the upper beam is tapered so that it has the same width at the tip but is wider at the clamping.
The taper ratio is such that the upper beam has an approximately $10\%$ higher fundamental natural frequency.

Every mechanical interconnection introduces some degree of uncertainty.
Thus, it may seem ideal to make the entire test rig out of one piece (monolithic design).
However, the impact interactions lead to wear so that the contacting parts must be exchangeable.
Moreover, the precise alignment of the impacting interfaces is much easier using fixtures than using a monolithic design.

In this work, we focus on analyzing the vibration behavior of the test rig.
MTU Aero Engines uses the test rig to characterize material properties.
The sensitive character of those investigations is the reason why the absolute dimensions of the test rig remain confidential so that only normalized quantities are reported.

The test rig is instrumented with accelerometers and laser-Doppler vibrometry (\fref{testrig}). 
More specifically, two accelerometers are placed on each beam (one at the tip and one at the middle) and three are placed onto the shaker platform (seven accelerometers in total).
This way, we determine the vibration response of the beams as well as the motion of the shaker platform.
The latter will be considered as an imposed base excitation in the simulation, as explained later in \sref{NLFR}.
Additionally, the contact activity is detected by applying a low electrical voltage and measuring the electrical resistance between the beams (beams are electrically isolated to the clamping).


\section{Modeling and simulation approach\label{sec:model}}
In the absence of impact interactions between the probes, it is assumed that the structure under test (beams with probes and clamping on shaker platform) behaves linearly.
In \ssref{linearmodel} we describe the finite element modeling of this underlying linear system and the linear model updating in \ssref{updating}.
Subsequently, we apply the modeling and simulation approach recently developed in \cite{Monjaraz.2021}.
This consists of reducing the model order using a massless-boundary approach, contact modeling and an appropriate time step integration technique, as described in \ssref{nonlinearmodel}.

\subsection{Modeling the system in the absence of impacts\label{sec:linearmodel}}
All parts of the test rig (\fref{testrig}) are modeled using solid finite elements, with the following exceptions:
Due to their small size, the seven accelerometers are simply modeled as point masses.
Also, the fixture components (bolts, nuts, probe holders) are not modeled as as individual bodies.
Instead, their mass is taken into account by increasing the density of the neighboring part accordingly.
More specifically, the mass of each probe's fixture components is accounted for in the respective probe,
and the mass of the \MOD{clamping fixture's} components is accounted for in the respective clamping block.
The elastic properties of the fixture components are not directly accounted for, which seems justified in view of the relatively low strain energy in the respective regions near resonance of the beams' fundamental bending mode.

As mentioned above, the imposed motion of the shaker platform will be considered as excitation.
Consequently, the lowest clamping block is rigidly tied to the ground at the interface to the shaker platform, and the excitation is modeled as distributed inertia loading.
The sought displacement field then corresponds to the displacement relative to the imposed motion \cite{Geradin.2014}.

The probes are made of a nickel alloy, the remaining solids (beams, clamping blocks) are made of steel.
Based on a mesh convergence study, with regard to the seven lowest-frequency normal vibration modes, the above mentioned solid parts are meshed using 156,000 finite elements.
Finite elements with quadratic shape functions are used everywhere except for the protrusion and their matching elements in the opposite probe indicated in \fref{rig_mesh}b.
This was modeled using hexahedral elements with linear shape functions because our nonlinear modeling approach only permits linear shape functions at the contact interface in its present form (\cf \ssref{nonlinearmodel}). 

\subsection{Linear model updating\label{sec:updating}}
A crucial prerequisite for any reasonable nonlinear dynamic validation is the good agreement of model and experiment with regard to the linear dynamics.
In the absence of impacts, the main uncertainty of the model is attributed to the fixtures.
In reality, the clamping does not act as a perfect tie constraint but it introduces finite stiffness and is an important source of damping.
The prediction of the effective stiffness and damping provided by bolted joints is a challenge itself \cite{Brake.2018}.
\MOD{
In addition to the bolted joints, a crepe tape was introduced between the beam and the clamping blocks (for electric isolation), whose properties are uncertain as well.
The focus of the present study is on the prediction of the nonlinear vibration response in the presence of impact interactions.
Consequently, the prediction of the behavior of the bolted joints and the crepe tape is considered as beyond the scope of this work.
Instead, a  state-of-the-art linear model updating approach is pursued, based on measurements in the regime without impacts.
}
\begin{table}[h!]
    \centering
    \fts{
\caption{
Modal frequencies and damping ratios in the absence of impact interactions.
$f_\mathrm{FE}$ corresponds to the updated finite element model.
The natural frequencies are normalized with respect to the second experimentally determined natural frequency.
The third mode was not identified experimentally.
}
        \label{tab:natfreqs}
        \begin{tabular}{|l|l||l|l|l|l|}
            \hline
            \textbf{mode no.} & \textbf{mode type} & $f_\mathrm{exp}/f_{\mathrm{exp},2}$ & $f_\mathrm{FE}/f_{\mathrm{exp},2}$ & \textbf{rel. error} & damping ratio $D_{\mathrm{exp}}$ \\ \hline
            1 & 1FL            & 0.918     & 0.922     & 0.44   \%    & 0.153\%   \\ \hline 
            2 & 1FU            & 1.000     & 1.020     & 1.96   \%    & 0.183\%   \\ \hline 
            3 & 1EL            & --        & 2.006     & --           & --        \\ \hline 
            4 & 1EU            & 2.400     & 2.445     & 1.87   \%    & 0.236\%   \\ \hline 
            5 & 1F clamp       & 3.509     & 3.552     & 1.24   \%    & 0.748\%   \\ \hline 
            6 & 2FL            & 5.763     & 5.579    & -3.21  \%    & 0.186\%   \\ \hline 
            7 & 2FU            & 5.950     & 5.851    & -1.65  \%    & 0.143\%   \\ \hline 
        \end{tabular}
    }
\end{table}
\myfig{htbp}{frflin}{Frequency response in the regime without impacts; $\omega_2=2\pi f_{\mathrm{exp},2}$}{0.96}
%

At the contact interfaces of the fixtures, linear-elastic layers are introduced.
All elastic layers representing the bolted joints at the clamping have the shape of a ring, with inner diameter equal to the bolt hole and outer diameter approximately equal to two times that diameter (\fref{rig_mesh}c).
These are assumed to have identical elastic properties.
Each beam is connected to its clamping block via four rectangular elastic layers.
Left and right layers, and upper and lower layers are assumed to have identical properties, respectively.
The elastic layers representing the connection between probes and beams have the shape of a disk.
Transversal isotropy is assumed for all above mentioned elastic layers so that each elastic layer has two individual properties, a normal stiffness and a tangential stiffness.
We have thus 7 different types of elastic layers giving rise to a number of 14 elastic properties in total.
These 14 parameters are treated as design variables of an optimization problem.
The objective is to minimize the deviation between the experimentally and numerically determined lowest natural frequencies as listed in \tref{natfreqs}.
Impact hammer modal testing, with force applied near the clamping and response measured at the laser point near the beam tip, is used to determine the natural frequencies experimentally.
The multi-objective optimization problem is solved using an evolutionary algorithm \cite{hadka12b}. 
Each objective function evaluation entails updating the elastic properties within the finite element model and subsequent numerical modal analysis.
%

The natural frequencies of the final updated finite element are confronted with the experimentally obtained ones in \tref{natfreqs}.
The nomenclature is as follows: \myquote{F} stands for flap-wise (vertical in \fref{testrig}-left) bending, \myquote{E} for edge-wise (horizontal) bending, \myquote{L} stands for lower arm, \myquote{U} for upper arm.
The third mode could not be identified experimentally because the signal-to-noise ratio was too low.
With the exception of mode six ($3.2\%$ deviation), the deviation is well below $2\%$.
This is regarded as sufficient for the purpose of the present study.
%

To further assess the quality of the updated linear model, frequency response functions are compared.
\fref{frflin}a and b depict the accelerance of the beam's tip with respect to impact hammer application point in vertical direction.
To obtain the results of the updated model, the experimentally identified modal damping ratios are adopted as reported in \tref{natfreqs}.
The missing damping ratios are set as $D_3=D_4=0.236\%$ and $D_{>7}=D_7=0.143\%$.
Based on the results, we conclude that the updated model is in good agreement with the experiment in the absence of impact interactions.
At the frequency of the 1FU mode, $\omega_2$, the lower beam participates only slightly.
Looking closely, one can infer that the updated model over-predicts the modal participation of the lower arm here.
While the ratio of the tip displacements is 115 in the experiment, the corresponding value is 47 in the updated model for the 1FU mode.
This deviation already present in the regime without impacts will be important for the interpretation of the results in the regime with impacts.

\subsection{Model order reduction, contact modeling, simulation\label{sec:nonlinearmodel}}
As detailed in the introduction, previous modeling and simulation endeavors truncated the elastodynamic model to one or a few normal modes, and approximated the effect of wave propagation / higher-frequency modes by \emph{calibrating} a restitution coefficient to experimental results in the \emph{regime with impacts}.
In contrast, we intend to model the impact interactions as predictive as reasonably possible, and rely only on experimental data gathered in the regime without impacts.
%

We apply the modeling and simulation approach proposed in \cite{Monjaraz.2021}.
This approach relies on the construction of a reduced-order model (ROM) using the MacNeal method, and a time step integration scheme accounting for the contact constraints.
In contrast to the more general formulation in \cite{Monjaraz.2021}, we use a simple node-to-node contact formulation in the present work.
This seems justified since the relative sliding distances remain small compared to the dimensions of the contact area.
Each contact area was meshed with $12\times16$ nodes and it was ensured that the meshes of either interface are conforming (\fref{rig_mesh}b).
Unilateral interaction is modeled in the normal direction and Coulomb's dry friction law is considered in the tangential plane.
The formulation and treatment of these set-valued contact laws is detailed later.

\subsubsection{ROM construction using MacNeal method}
To reduce the mathematical model order of the finite element model, we use MacNeal's component mode synthesis method.
This method distinguishes between boundary coordinates, which are retained, and inner coordinates, which are reduced.
It is common practice to define the nodal coordinates on both sides of the contact interface as boundary coordinates.
As described above, constant node pairing is assumed.
This allows us to define an invariant transform from nodal displacements on either side of the contact interface to relative and absolute displacements.
We retain only the relative displacements, $\mm g$, as boundary coordinates to further reduce the model order.
These are defined in the local contact coordinate system, \ie, $\mm g = [\mm g_1;\ldots;\mm g_{\ncon}]$ with the number of contact node pairs $\ncon$ and $\mm g_k = [g_{\mathrm n,k};\mm g_{\mathrm t,k}]$ where $g_{\mathrm n,k}$ is the normal contact gap and $\mm g_{\mathrm t,k}$ the two-dimensional tangential relative displacement of node pair $k$, respectively.
The remaining coordinates, $\mm q_{\mathrm i}$, contain all independent\footnote{Here, \emph{independent} refers to those degrees of freedom that are neither constrained (interface to the shaker platform) nor part of the dependent contact interface.} nodal degrees of freedom.
%

The coordinates $\mm g$, $\mm q_{\mathrm i}$ are approximated as a linear combination of a set of component modes, 
\e{
	\qq = \vector{\mm g\\ \qqi} \simeq  \mm R \vector{\mm g \\ \mm\eta} \fp
}{MacNeal}
The component modes are the columns of the matrix $\mm R$.
In the MacNeal method, the component modes are a subset of free-interface normal modes (associated with the modal coordinates $\mm\eta$) and residual flexibility attachment modes \cite{macn1971}.
\MOD{
As the interface is at the end of the beam, the free-interface normal modes are essentially cantilevered beam modes (flap-wise/edge-wise bending, torsion modes).
The residual flexibility attachment modes are static deformation shapes in response to unit forces applied to the different degrees of freedom of the nodes within the contact interface, where the contribution of the retained free-interface normal modes has been subtracted.
}
The component modes are determined from the mass and stiffness matrices of the finite element model.
We exported the FE matrices from \CalculiX, applied the above described coordinate transform, and carried out the model order reduction in \Julia.
The MacNeal method yields reduced mass and stiffness matrices, which are defined, along with $\mm R$, \eg in \cite{Monjaraz.2021}.
An important property of the MacNeal method is that the static flexibility with respect to loads applied at the boundary (coordinates $\mm g$) is described as accurately as the parent finite element model \cite{macn1971}.
To properly model the elastodynamics, the retained normal modes should cover the relevant frequency range of the response.
The 200 lowest-frequency free-interface normal modes were retained.
The highest considered modal frequency is $180 f_{\mathrm{exp},2}$, which is much higher than the highest frequency identified during the linear experimental modal analysis (\cf \fref{frflin}, \tref{natfreqs}).
A stricter truncation leads to a poorer representation of the wave propagation and its effect on the impact interactions.
With a further increase of the number of normal modes, on the other hand, the results depicted throughout this work did not change significantly.
%

It is well-known that the MacNeal method leads to a singular mass matrix, where no inertia is associated with the boundary coordinates.
In linear dynamics, this is commonly viewed as deficiency.
In \cite{Monjaraz.2021} it was noted, for the first time, that the resulting massless boundary makes the method attractive for dynamic contact problems.
Standard finite elements associate a finite mass with the degrees of freedom at the contact boundary.
The mass-carrying boundary is known to cause spurious high-frequency oscillations and/or poor energy conservation properties \cite{Khenous.2008}.
The proposed massless-boundary approach overcomes these problems.
It should be remarked that alternative approaches exist for mass redistribution, including the modification of the location of the integration points \cite{Hager.2008,Hager.2009} or the adjustment of the shape functions \cite{Renard.2010,Tkachuk.2013}.
Rather than redistributing the mass on finite element level, we achieve the massless boundary via component mode synthesis in this work.
This is convenient because this allows us to use a conventional finite element tool to construct the parent model.

\subsubsection{Semi-explicit time step integration and contact treatment}
\MOD{As the vibrations in the considered problem setting turned out to be non-periodic}, many computationally efficient methods such as Harmonic Balance in conjunction with numerical continuation \cite{Krack.2019} are not applicable, and numerical time step integration has to be used instead.
We use Verlet's scheme, which is second-order accurate (fixed contact conditions), time-reversible and has favorable energy conservation properties (symplectic integrator).
We use the notation $\mm g^j$ \MOD{for the approximation of} $ \mm g(t^j)$, where $t^j$ is the $j$-th time level, and we assume a fixed time step $\Delta t$.
The stepping scheme is written in leapfrog form; \ie, the grid of the velocities is shifted by a half time step with respect to that of the displacements.
The resulting equations can be summarized as:
\ea{
\mm K_{\mathrm{gg}}\mm g^j + \mm K_{\mathrm g\eta}\mm\eta^j - \mm \lambda^j &=& \mm 0 \fk \label{eq:algebraic} \\
\mm g^j &=& \mm g^{j-1} + \mm \gamma^{j-\frac12}\Delta t \fk \label{eq:boundaryStep} \\
\frac{\dot{\mm\eta}^{j+\frac12} - \dot{\mm \eta}^{j-\frac12}}{\Delta t} + \mm D_{\eta\eta}\frac{\dot{\mm\eta}^{j+\frac12} + \dot{\mm \eta}^{j-\frac12}}{2} + \mm K_{\mathrm g\eta}^{\mathrm T} \mm g^j + \mm K_{\eta\eta}\mm\eta^j &=& - \mm \beta \ddot{q}_{\mathrm{base}}^j \fk \label{eq:differential} \\
\mm\eta^{j+1} &=& \mm\eta^j + \dot{\mm\eta}^{j+\frac12}\Delta t \fk \label{eq:explicit}
}
\MOD{
\eref{differential} expresses the dynamic force balance within the solid bodies, while \eref{algebraic} expresses the static force balance at the contact boundary.
Compared to the case of a mass-carrying boundary, the differential index of the mathematical problem reduces from 3 to 1, which substantially reduces the numerical contact oscillations.
}
$\mm K_{\mathrm{gg}}$, $\mm K_{\mathrm g\eta}$ and $\mm K_{\eta\eta}$ are sub-matrices of the reduced stiffness matrix.
The symmetric coefficient matrix $\mm D_{\eta\eta}$ accounts for the experimentally identified modal damping ratios.
The MacNeal method yields a very simple reduced mass matrix (identity matrix with respect to $\mm\eta$, otherwise zeros), and this is exploited in \erefs{algebraic} and \erefo{differential}.
The vector of gap velocities, $\mm\gamma$, is simply the time derivative of $\mm g$.
$\mm\lambda$ is the vector of contact forces and it is related to $\mm\gamma$, $\mm g$ via contact laws, as described later.
The term $\mm\beta\ddot{q}_{\mathrm{base}}$ models the distributed inertia loading, where $\ddot{q}_{\mathrm{base}}$ is the base acceleration and $\mm\beta = \mm\Phi^\mathrm T_{\mathrm i}\,\mm M_{\mathrm{ii}}\mm b$, with the finite element mass matrix restricted to the inner coordinates, $\mm M_{\mathrm{ii}}$, the matrix of free-interface normal modes restricted to the inner coordinates, $\mm\Phi_{\mathrm i}$, and a Boolean vector, $\mm b$, with entry one if the corresponding coordinate is equally oriented as the base motion and zero if it is orthogonal.
%

The algorithm to solve \erefs{algebraic}-\erefo{explicit} is as follows:
For given $\mm\eta^j$ and $\mm g^{j-1}$, \eref{boundaryStep} is substituted into \eref{algebraic}, which is then solved \emph{implicitly} together with the contact laws for $\mm\gamma^{j-\frac12}$ and $\mm\lambda^j$.
The implicit contact treatment is described in the next paragraph.
Then, $\mm g^j$ is updated using \eref{boundaryStep}.
Subsequently, \eref{differential} is solved \emph{explicitly} for $\dot{\mm\eta}^{j+\frac12}$.
Finally, $\mm\eta^{j+1}$ is updated using \eref{explicit}.
The process repeats until the end of the simulation is reached.
%

Unilateral interaction (Signorini conditions) is considered in the normal contact direction.
Coulomb's law of dry friction is considered in the tangential contact plane.
These can be written as complementary inequality constraints, or equivalently, as inclusions $- \mm \gamma^{j-\frac12} \in \mathcal N_{\mathcal C}\left(\mm\lambda^j\right)$, into the normal cone $\mathcal N_{\mathcal C}$, to the admissible set $\mathcal C$.
$\mm\lambda$ is sorted analogous to $\mm g$.
For the above mentioned contact laws, the admissible set for each active contact (for which $g_{\mathrm n,k}\leq 0$) is $\mathcal C_k = \mathbb R_0^+ \times \mathcal D\left(\mu\lambda_{\mathrm n,k}\right)$, where $\mathcal D\left(r\right)$ denotes the planar disk of radius $r$.
The friction coefficient is set as $\mu=0.4$ and applies to both static and dynamic friction.
In the proposed algorithm, \erefs{algebraic}-\erefo{boundaryStep} are used to obtain an expression of the form $\mm \gamma^{j-\frac12} = \mm G\mm\lambda^j+\mm c$.
Subsequently, the implicit algebraic inclusion $-\left(\mm G\mm\lambda^j+\mm c\right)\in\mathcal N_{\mathcal C}\left(\mm\lambda^j\right)$ is solved using an augmented Lagrangian technique in conjunction with a projected Jacobi relaxation scheme.
Details on the contact treatment, including initialization, active set strategy and the solution of the inclusions are given in \cite{Monjaraz.2021}.
%

\MOD{
A popular alternative to the proposed contact modeling approach is penalty regularization.
In the one-dimensional case, both approaches lead to a mathematically equivalent problem.
However, the delicate selection of the penalty stiffness cannot be reasoned by purely physical arguments.
In contrast, the stiffness $\mm K_{\mathrm{gg}}$ in \eref{algebraic} is consistently derived from geometry and material properties.
Furthermore, the mathematical equivalence is lost in the two- and three-dimensional case, because there is elastic coupling between the boundary coordinates in the proposed approach, in contrast to penalty regularization.
}
\section{Nonlinear vibration response under impact interactions\label{sec:NLFR}}
We finally analyze the system's vibration behavior in the presence of impact interactions.
Forward and backward stepped sine tests were carried out in the frequency range near the fundamental vertical bending frequency of the upper arm (which is the second-lowest natural frequency of the system, \cf \tref{natfreqs}).
If the excitation was provided by a small shaker via a stinger, the severe impact nonlinearity would be expected to cause strong exciter-structure interaction.
That is why we decided to provide the excitation via a large shaker in the form of base motion.
As the total moving mass (shaker platform, clamping blocks, beams) is relatively large compared to the modal mass, a relatively weak exciter-structure interaction is expected.
To estimate the relevance of exciter-structure interaction and assess the quality of the excitation, we monitored the base motion using the sensors indicated in \fref{testrig}.
By integrating the velocity and acceleration signals, we obtain the base displacement.
It was found that the base displacement is indeed dominated by the fundamental harmonic (for harmonic voltage input to the shaker).
As expected under severe impacts, the base displacement also contains a small non-periodic contribution.
More importantly, the well-known resonant force drop was encountered \cite{Tomlinson.1979}, both in the absence and in the presence of impacts.
Therefore, we used feedback control to maintain a constant fundamental harmonic magnitude of the base displacement throughout the frequency steps. 
The fundamental harmonic component was determined using synchronous demodulation as in \cite{Schwarz.2019}.
%

We tested four different harmonic base displacement levels, $\hat q_{\mathrm{base}} \in [0.5;0.8;2;6]\cdot 10^{-5}\ell$, where $\ell$ is the beams' length.
For brevity, results are only depicted for the highest two excitation levels for which the most severe impact interactions occur.
Substantial wear of the contact interface was noticed, especially during the forward and backward stepping at the highest excitation level.
The prediction of the material loss is considered beyond the scope of the present work, which focuses on the prediction of the nonlinear vibration behavior.
Instead, the actual clearance was experimentally identified and updated in the model.
Unfortunately, the clearance was not directly measured between the stepped sine tests.
Thus, the clearance was estimated based on the measured relative displacement at the beams' tip just before the first impact occurred during a given stepped sine test.
The clearance was then kept constant in the model during the entire stepped sine analysis.
The order and specifics of the reported stepped sine tests, along with the estimated clearances are listed in \tref{clearance}.
%
\begin{table}[h!]
    \centering
    \caption{Order of stepped sine tests and estimated clearance before respective test}
    \label{tab:clearance}
    \fts{
        \begin{tabular}{|l||l|l|l|}
            \hline
            \textbf{Test number} & \textbf{Level in $10^{-5}\ell$} & \textbf{Stepping direction} & \textbf{estimated clearance before test in $10^{-3}\ell$} \\ \hline
            1 & 2.0 & down  & 2.33 \\ \hline
            2 & 2.0 & up    & 2.33 \\ \hline
            3 & 6.0 & up    & 2.41 \\ \hline
            4 & 6.0 & down  & 2.78 \\ \hline
        \end{tabular}
    }
\end{table}
%
%

We also carried out a stepped sine test at a level just before impacts occur.
From the results, we identified a modal damping ratio.
This damping ratio was only $57\%$ of that identified by impact hammer modal testing.
We attribute this observation to nonlinear damping behavior of the clamping.
Additionally, we noted that the corresponding amplitude-frequency curve exhibited a slight skewness typical of dry friction damping.
This further supports our hypothesis of slightly nonlinear clamping behavior.
We updated the corresponding modal damping ratio to the value identified from the stepped sine test without impacts, keeping the remaining damping ratios unchanged.

\subsection{Overview of frequency response analysis}
An overview of the frequency response results is given in \fref{plot_rms_5_15fexscal}.
\MOD{
As the response is non-periodic, the root-mean-square (RMS) value of the beams' tip velocity in vertical direction is used as amplitude measure.
Based on results with increased wait time (before the time series is recorded) and increased recording time at each frequency step, it was ensured that the transients have sufficiently decreased and the quantities depicted throughout this paper have converged.
}
The quantity $\hat u_{\mathrm{tip}}^{\mathrm{rms}}$ is normalized by the angular excitation frequency $\Omega$ and the base displacement level.
The excitation frequency is normalized by the linear natural frequency $\omega_2=2\pi f_{\mathrm{exp},2}$ of the resonant mode (fundamental bending mode of upper arm).
The results for all four tests listed in \tref{clearance} are depicted, \ie, for both levels and both stepping directions.
The response of both the upper and the lower arm is shown.
%
\begin{figure}[htbp]
    \centering
    \includegraphics[width=0.99\textwidth]{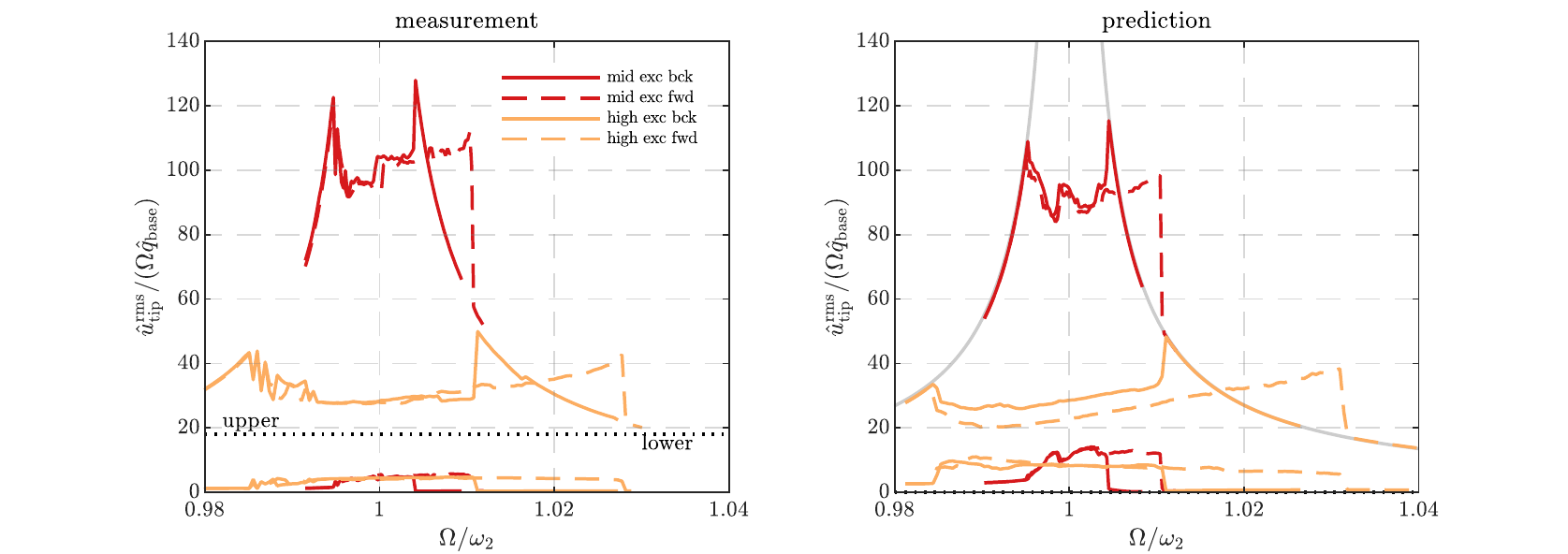}
    \caption{Frequency response in terms of root-mean-square (RMS) value of tip velocity}
    \label{fig:plot_rms_5_15fexscal}
\end{figure}
%
%

Overall, prediction and measurement are in very good agreement, especially in the light of the severe impact interactions and the resulting strongly nonlinear response.
In particular, the frequency span in which impacts occur is similar and the root-mean-square response levels are matching well, for both forward and backward stepped sines.
%

In the linear case without impacts, the lower arm almost does not vibrate.
This is because of the fact that the lower arm's fundamental bending mode is about $10\%$ smaller than the considered resonance frequency.
In the presence of impacts, nonlinear energy transfer to the lower arm occurs.
Consequently, the response of the lower arm almost instantaneously increases upon the first impacts.
The model over-predicts this effect, consistent with the deviation observed already for the underlying linear model (in the regime without impacts), \cf \ssref{updating}.
%

Due to the hardening effect induced by closing contacts, co-existing vibration states are expected in a certain frequency band.
Indeed, the results of forward and backward stepping deviate, and a pronounced jump occurs at the transition from the regime with to without impacts during the forward stepping sequences.
In addition, at the higher excitation level, forward and backward stepping deviate also at lower excitation frequencies, before the regime with no impacts is reached and a unique response is obtained.
The co-existence of vibration states in that regime is less expected and is a further indicator of the strong nonlinearity.
The distinction between forward and backward stepping in that regime is better visible in the prediction.
Still, in the experiment, the \myquote{zig-zag}-behavior near $\Omega/\omega_2=0.99$ also suggests jumps between two co-existing branches of vibration states.
\begin{figure}[htbp]
    \centering
    \includegraphics[width=0.99\textwidth]{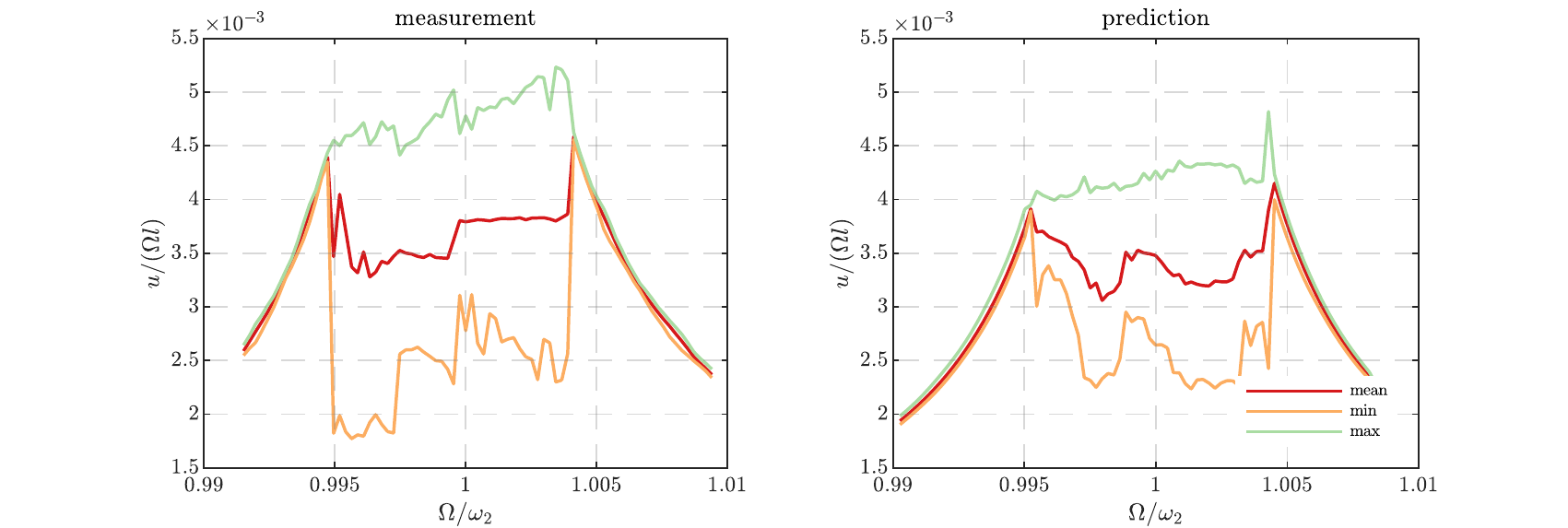}
    \caption{Frequency response in terms of average, minimum and maximum amplitude of tip velocity of upper arm for lower excitation level}
    \label{fig:plot_minmax_05}
\end{figure}
\begin{figure}[htbp]
    \centering
    \includegraphics[width=0.99\textwidth]{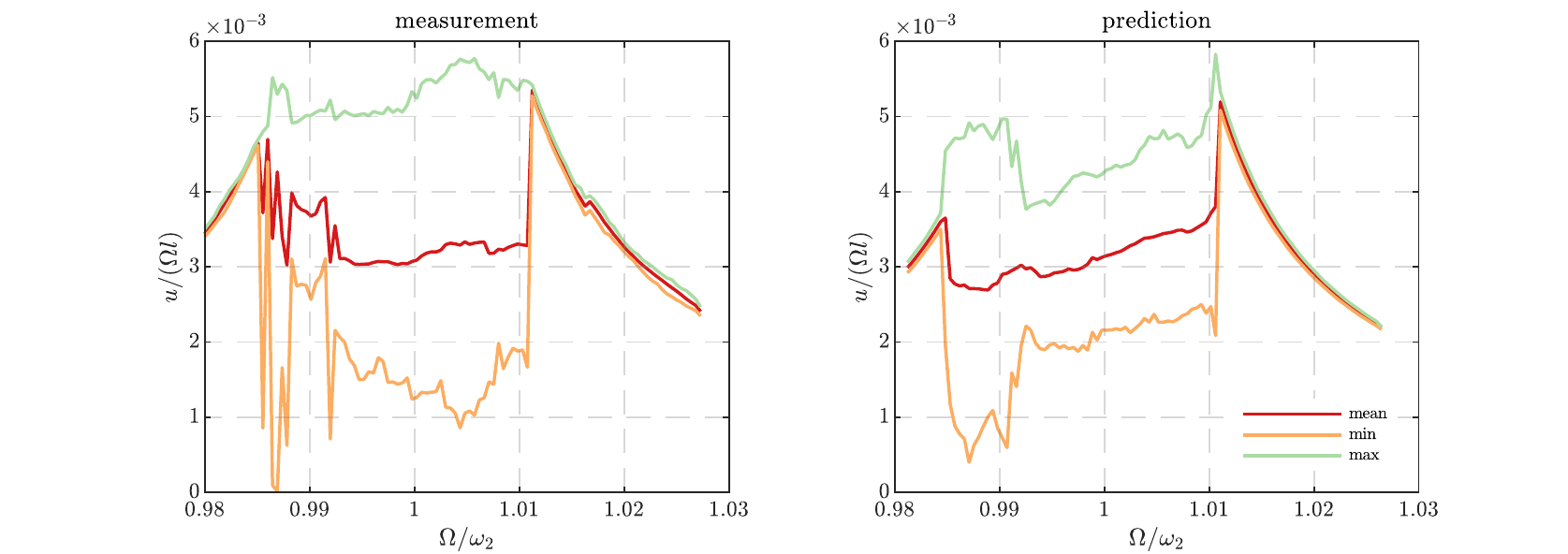}
    \caption{Frequency response in terms of average, minimum and maximum amplitude of tip velocity of upper arm for higher excitation level}
    \label{fig:plot_minmax_15}
\end{figure}
%
%

Interestingly, the root-mean-square value of the response is lower in the regime with impacts.
\frefs{plot_minmax_05} and \frefo{plot_minmax_15} depict how minimum, maximum and mean amplitudes evolve with frequency (only for backward frequency stepping).
To obtain these values, the zero-to-peak amplitudes in each excitation period are determined, and the extreme and average values are calculated from this set.
The results clearly show that the maximum amplitude indeed reaches its largest value in the regime with impacts, as expected.
Moreover, it can be inferred that the response is far from being one-periodic in the regime with impacts.
In fact, the response is strongly modulated there, as further analyzed in the next subsection.
Again, prediction and experiment are in good agreement.
%

In the simulation, the time step was varied in a large range.
A number of $1,\xspace000$ time steps per excitation period was found to be a good compromise between accuracy and computational cost.
This way, the highest natural frequency is sampled by at least $5$ time levels, and this is in the order of magnitude of the maximum stable time step for explicit methods (open contact conditions).
The wall time to obtain the nonlinear simulation results for all figures presented in this work was 21 hours using a single Intel Xeon Gold 6134 processor.

\subsection{Detailed analysis for a representative frequency}
\label{sec:res_detailed}
We further analyze the nonlinear response for a certain, representative excitation frequency, $\Omega=\omega_2$.
From the time history of the beams' tip velocity depicted in \frefs{velsignal_mid} and \frefo{velsignal_high}, one can clearly infer the strongly modulated nature of the response.
The modulation is more pronounced for the higher excitation level.
Prediction and experiment are in qualitative and reasonable quantitative agreement, especially in the light of the dependence on initial conditions inherent to non-periodic (most likely chaotic) oscillations.
The over-prediction of the lower beam's response is in line with the previous observations.
\begin{figure}[htbp]
    \centering
    \begin{subfigure}[b]{0.45\textwidth}
        \centering
        \includegraphics[scale=1]{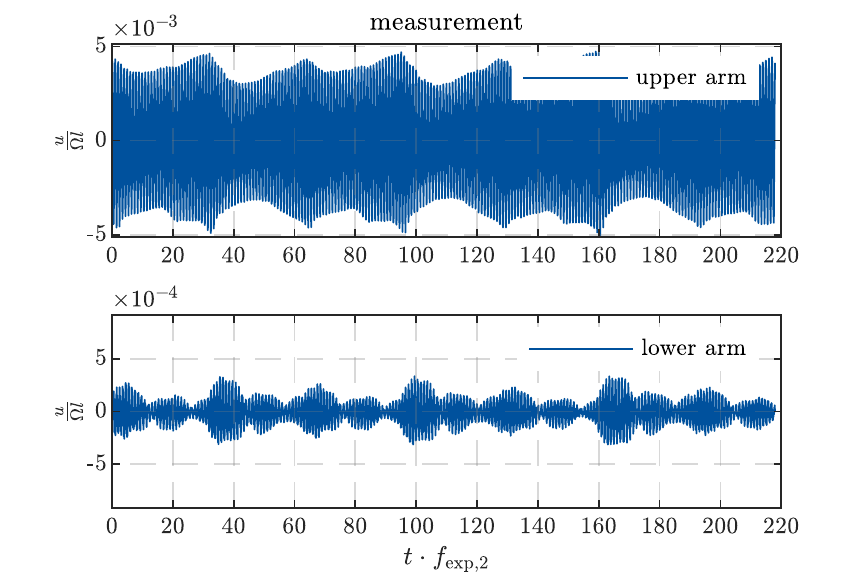}
    \end{subfigure}
    \hfill
    \begin{subfigure}[b]{0.45\textwidth}
        \centering
        \includegraphics[scale=1]{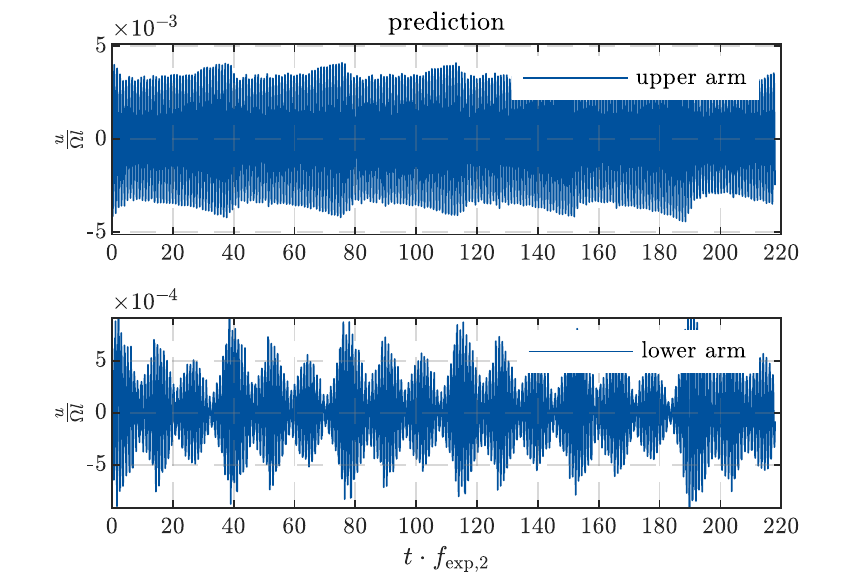}
    \end{subfigure}
    \caption{Time evolution of a typical steady-state section of the tip velocity for $\Omega / \omega_2 = 1$, lower excitation level}
    \label{fig:velsignal_mid}
\end{figure}
\begin{figure}[htbp]
    \centering
    \begin{subfigure}[b]{0.45\textwidth}
        \centering
        \includegraphics[scale=1]{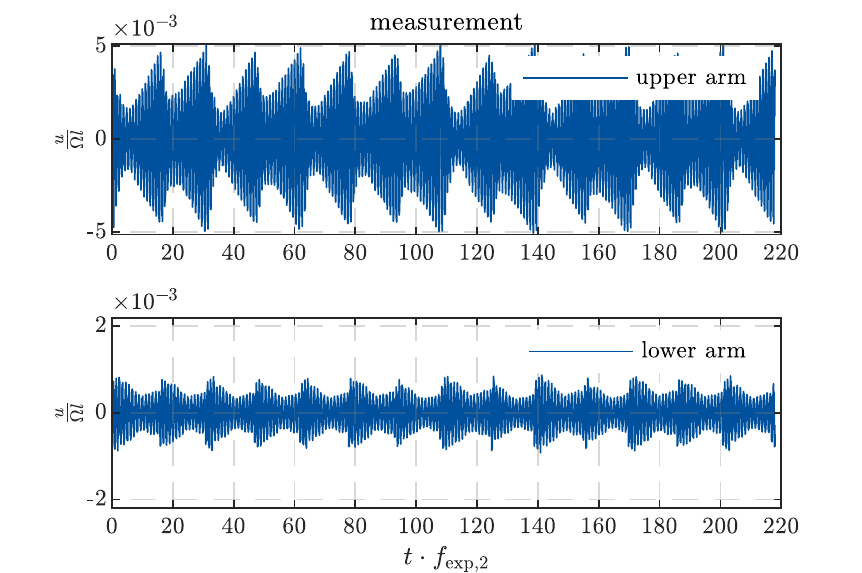}
    \end{subfigure}
    \hfill
    \begin{subfigure}[b]{0.45\textwidth}
        \centering
        \includegraphics[scale=1]{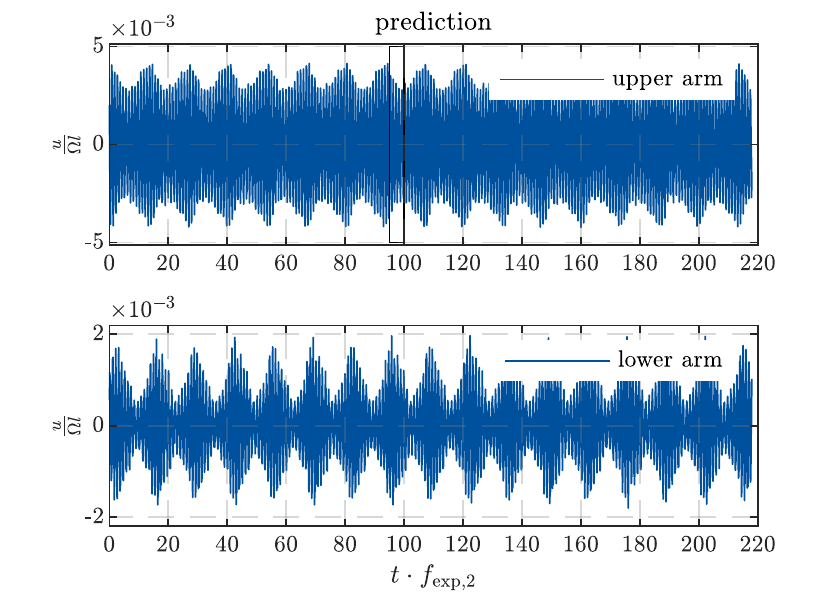}
    \end{subfigure}
        \hfill
        \begin{subfigure}[b]{0.45\textwidth}
        \centering
        \includegraphics[scale=1]{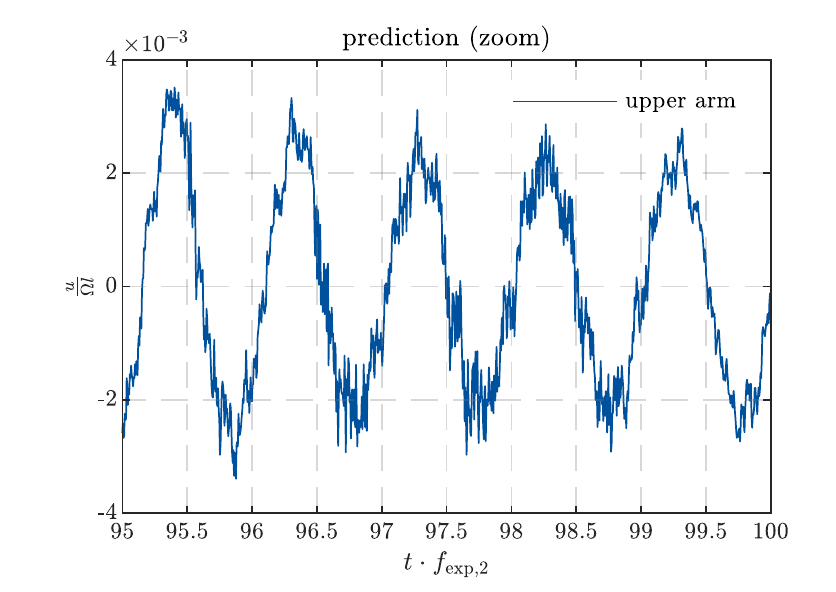}
    \end{subfigure}
    \caption{Time evolution of a typical steady-state section of the tip velocity for $\Omega / \omega_2 = 1$, higher excitation level. \MOD{Below; zoom of the prediction at the upper arm.}}
    \label{fig:velsignal_high}
\end{figure}

In \fref{contactactivity}, the contact activity is illustrated for the higher excitation level.
Recall that the contact activity was determined by measuring the electrical resistance between the otherwise isolated beams.
In the simulation, the contact is considered as active at a certain time level if the gap is closed at one or more node pairs.
Apparently, the strongly modulated response is associated with phases in which no impacts occur and phases with multiple consecutive impacts.
For the given parameters, clusters of five impacts occur recurrently.
The predicted contact pattern is in excellent agreement with the experiment.
\MOD{In addition, the predicted contact forces are shown in \fref{contactactivity}.}
\begin{figure}[htbp]
    \centering
    \begin{subfigure}[b]{0.45\textwidth}
        \centering
        \includegraphics[scale=1]{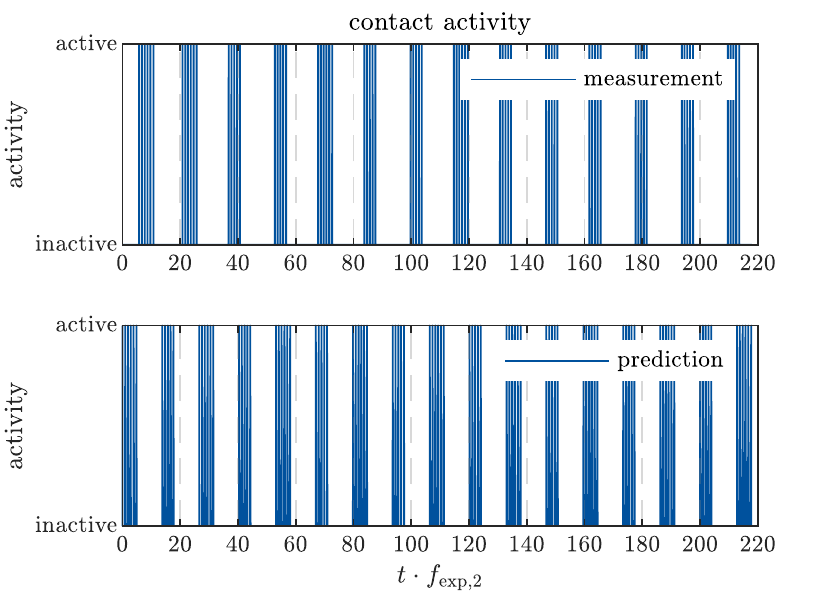}
    \end{subfigure}
    \hfill
    \begin{subfigure}[b]{0.45\textwidth}
        \centering
        \includegraphics[scale=1]{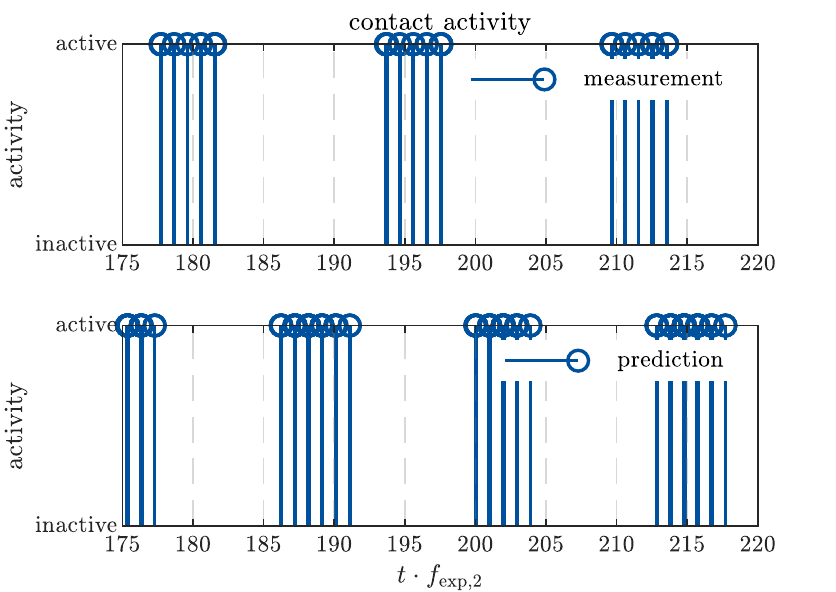}
    \end{subfigure}
        \hfill
    \begin{subfigure}[b]{0.45\textwidth}
        \centering
        \includegraphics[scale=1]{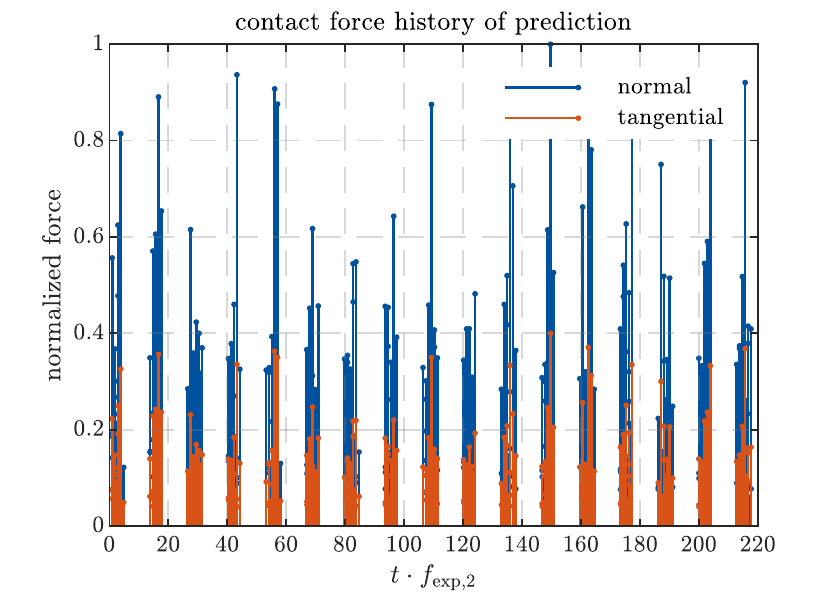}
    \end{subfigure}
    \caption{Time evolution of a typical steady-state section of the contact activity for $\Omega / \omega_2 = 1$, higher excitation level; right plots are zooms into the late time span of the left plots;
    \MOD{time evolution of normal and tangential contact forces (integral value over entire contact interface), normalized by maximum within displayed range.}
    }
    \label{fig:contactactivity}
\end{figure}
%
%

Next, a time-frequency analysis of the normalized velocity $u/(\Omega\ell)$ is carried out using the wavelet transform as described in \cite{carassale.2018}.
The results are depicted in \fref{wavelets}.
The color scale represents the logarithm (to the base 10).
One can deduce a strong contribution of multiple frequencies.
In particular, higher harmonics of the excitation frequency appear.
The frequencies of first and second vertical bending modes are close to a ratio of $1:6$ (\tref{natfreqs}).
Thus, the $6$th harmonic of the excitation, generated by the strong impact nonlinearity, may engage into resonance with the second bending modes.
This explains the strong participation of the corresponding frequency according to \fref{wavelets}.
Again, prediction and experiment are in very good agreement.
\begin{figure}[htbp]
    \centering
    \begin{subfigure}[b]{0.45\textwidth}
        \centering
        \includegraphics[scale=1]{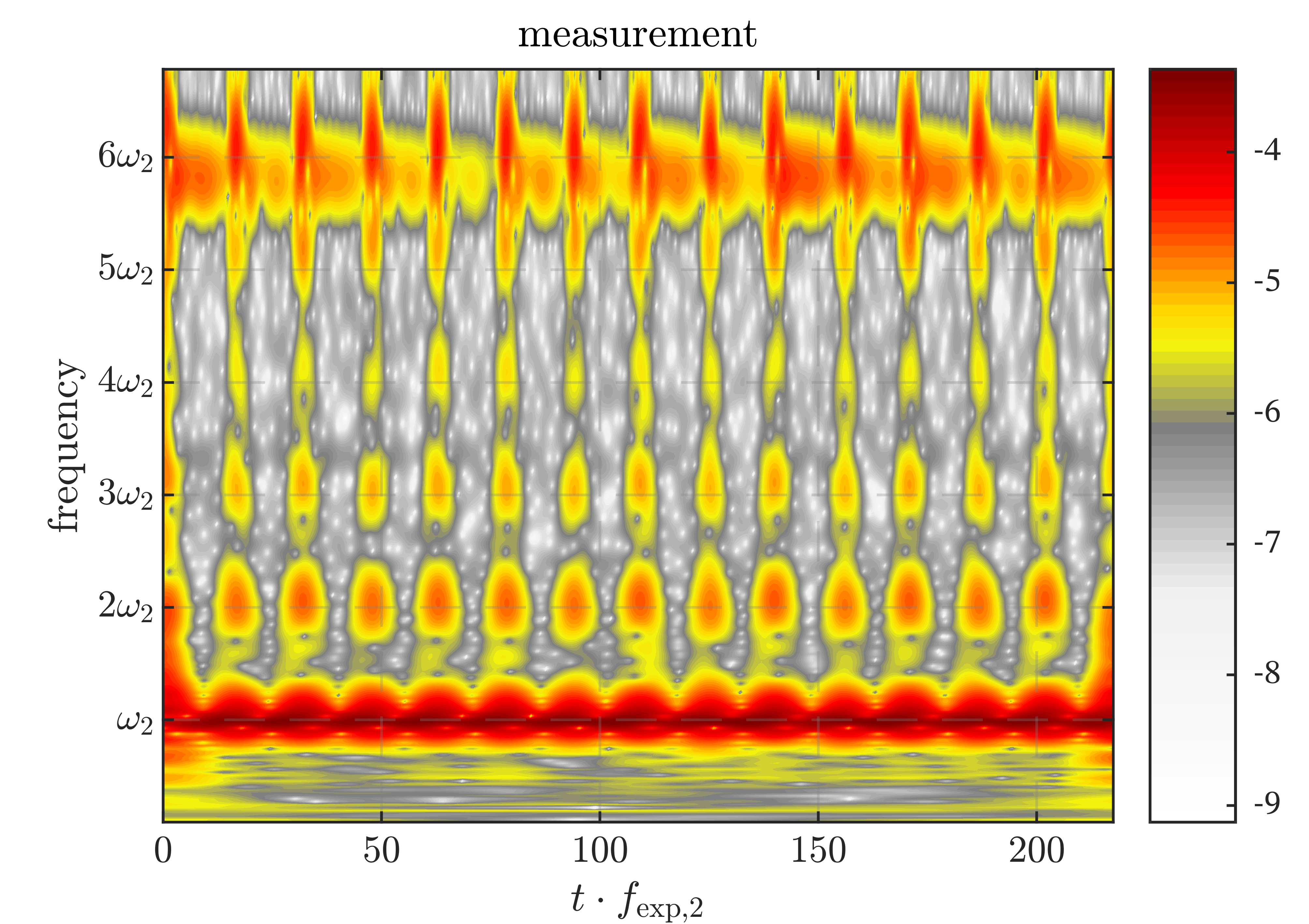}
    \end{subfigure}
    \hfill
    \begin{subfigure}[b]{0.45\textwidth}
        \centering
        \includegraphics[scale=1]{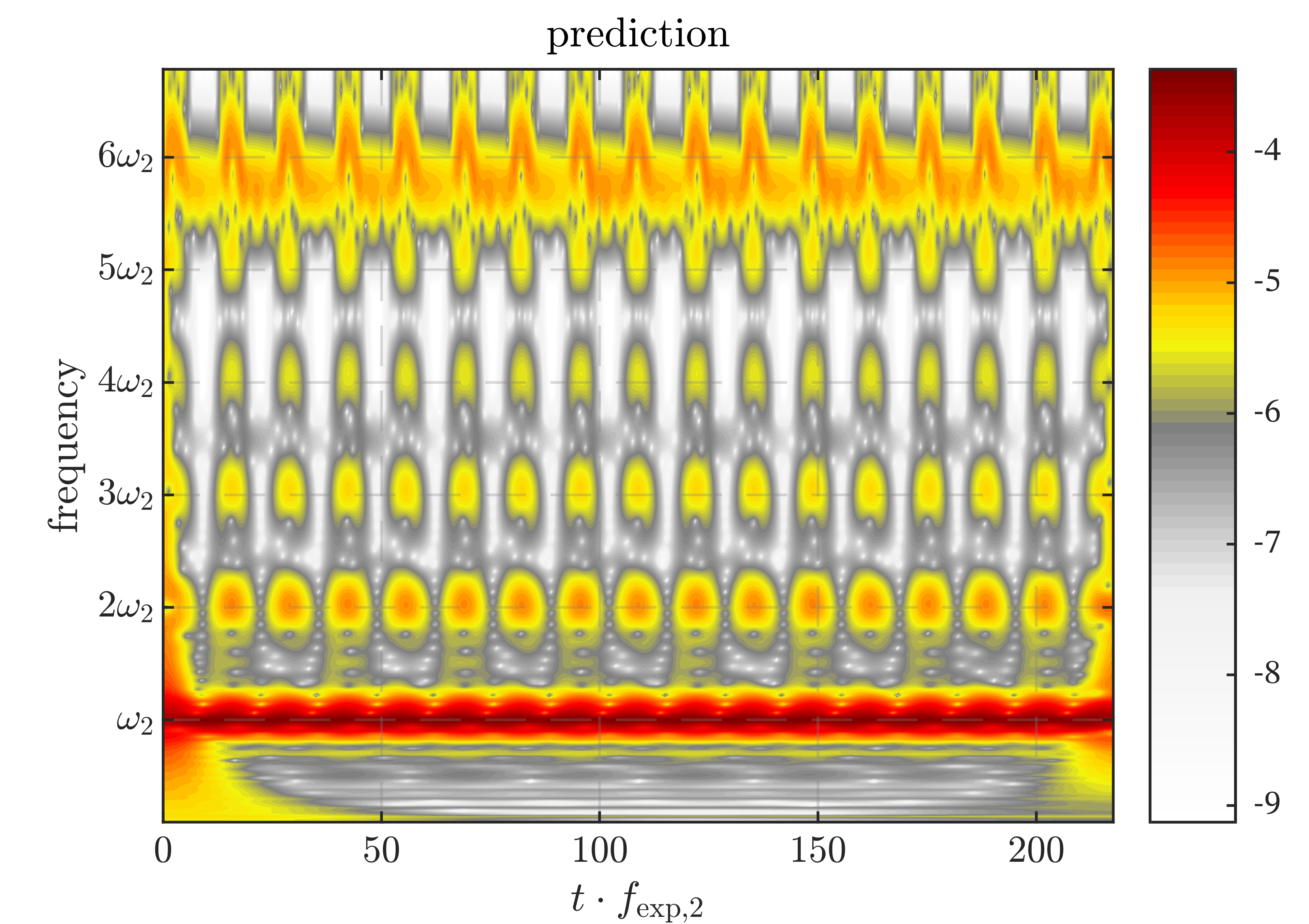}
    \end{subfigure}
    \caption{Frequency content over time of a typical steady-state section of the upper beam's tip velocity for $\Omega / \omega_2 = 1$, higher excitation level. The color scale is logarithmic.}
    \label{fig:wavelets}
\end{figure}
%

\subsection{Convergence analysis}
\MOD{
The convergence with the number of time steps per excitation period is depicted in \fref{convergence_time}.
As response quantity, the upper beam's tip velocity obtained at frequency $\Omega=\omega_2$ for the higher excitation level is used (as depicted in \fref{velsignal_high}-top-left).
Due to the chaotic nature of the response, the long-term time evolution is unpredictable.
Instead, we provide the same initial conditions throughout all simulations and consider only the time span of the first five excitation periods.
We use the relative root-mean-square error,
\begin{equation}
    	\varepsilon_\mathrm{RMS} =
    	\sqrt{
    		\frac{\sum_{k}  \left|  u^k_{\mathrm{tip}} -
    	                      u^k_{\mathrm{tip,ref}} \right|^2 }
    	    	   { \sum_k \left| u^k_{\mathrm{tip,ref}} \right|^2 }
    	    }
    	    	   .
\end{equation}
As reference, we use the result obtained for the finest time step with 128,000 time levels per excitation period.
The estimated error decreases monotonously until it reaches a round-off plateau.
Recall that we selected a number of 1,000 time levels per excitation period for all results shown before.
A further increase of the time step size led to divergence.
As the results in \fref{convergence_time}a suggest, a further decrease of the step size does not lead to significantly different results.
Indeed, the root-mean-square value of the velocity, which was illustrated \eg in \fref{plot_rms_5_15fexscal}, changes by less than $1\%$ compared to the reference (smallest considered time step).
Of course, the convergence behavior depends on the considered response quantity.
More specifically, accelerations and forces are expected to converge less well and displacements are expected to converge better than the considered velocities.
}

\MOD{
Finally, the convergence with respect to the number of free-interface normal modes retained in the MacNeal reduction basis is analyzed.
This is done largely analogous to what is described above.
The results are depicted in \fref{convergence_modal}.
As reference, the results obtained for 800 normal modes is used.
Recall that we retained 200 normal modes to produce the results depicted before.
The error decreases not strictly monotonously but rather in a stepped way.
This is due to the fact that not all (types of) modes contribute significantly in the response (dominated by vertical bending).
From 80 normal modes on, the root-mean-square value of the velocity changes by less than $1\%$, and the time evolution does not significantly change anymore.
It should be remarked that the initial conditions are slightly different between \fref{convergence_time} and \frefo{convergence_modal}.
More specifically, the stepped-sine excitation was carried out again with 800 modes in order to obtain the initial values of the higher modes to generate \fref{convergence_modal}.
 because the (nonzero) initial values of the higher modes had to be consistent with the stepped-sine procedure (and are different from zero).
}
\begin{figure}[htbp]
    \centering
    \begin{subfigure}[b]{0.45\textwidth}
        \centering
        \includegraphics[scale=1]{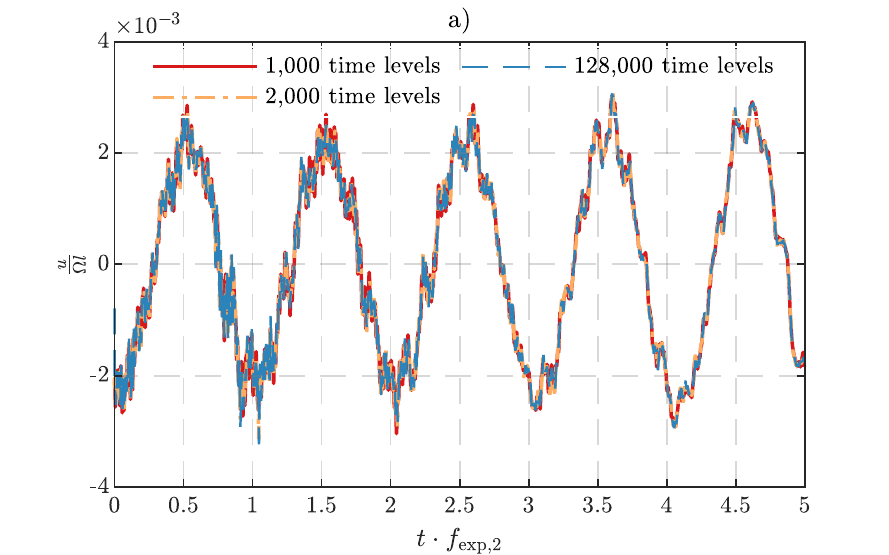}
    \end{subfigure}
    \hfill
    \begin{subfigure}[b]{0.45\textwidth}
        \centering
        \includegraphics[scale=1]{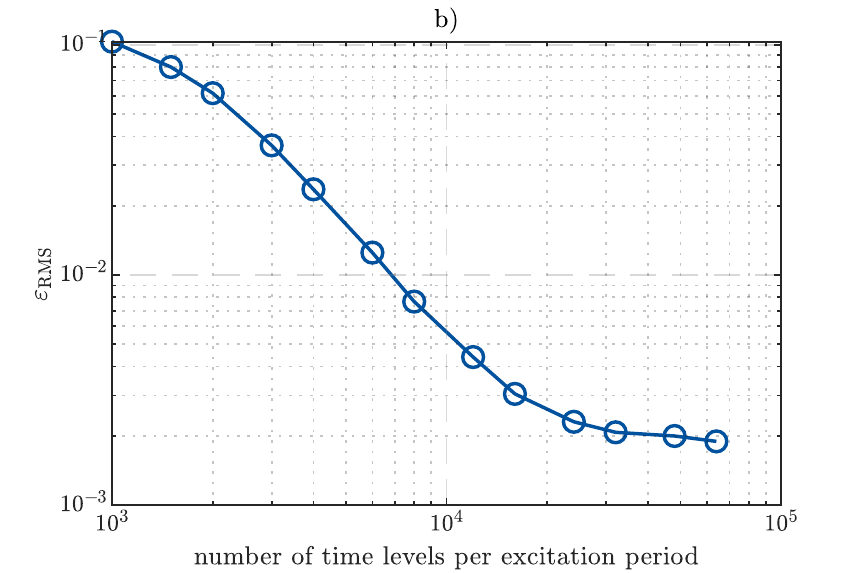}
    \end{subfigure}
    \caption{
    \MOD{Convergence with respect to time step: (a) representative time interval of the upper beam's tip velocity; (b) relative root-mean-square error; the representative time interval corresponds to the first five periods illustrated in \fref{velsignal_high}-top-left}
    }
    \label{fig:convergence_time}
\end{figure}
%
%
\begin{figure}[htbp]
    \centering
    \begin{subfigure}[b]{0.45\textwidth}
        \centering
        \includegraphics[scale=1]{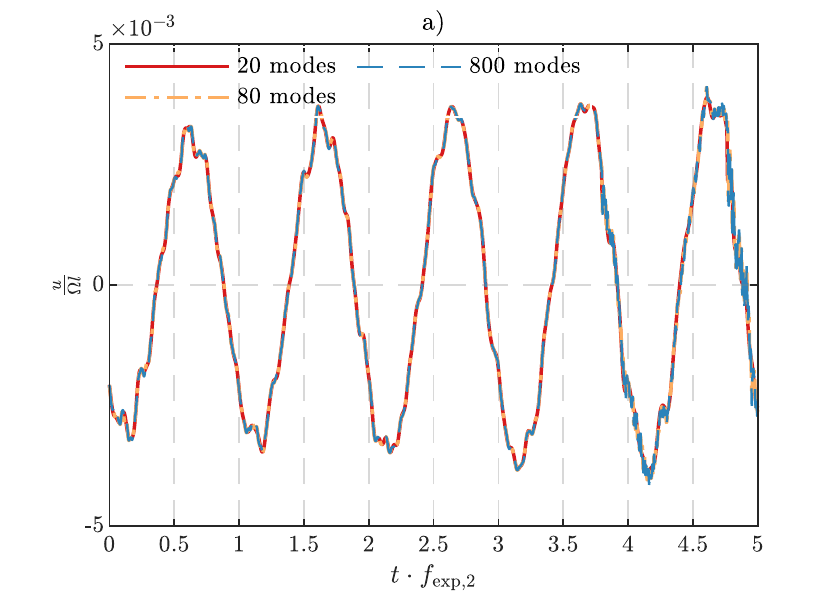}
    \end{subfigure}
    \hfill
    \begin{subfigure}[b]{0.45\textwidth}
        \centering
        \includegraphics[scale=1]{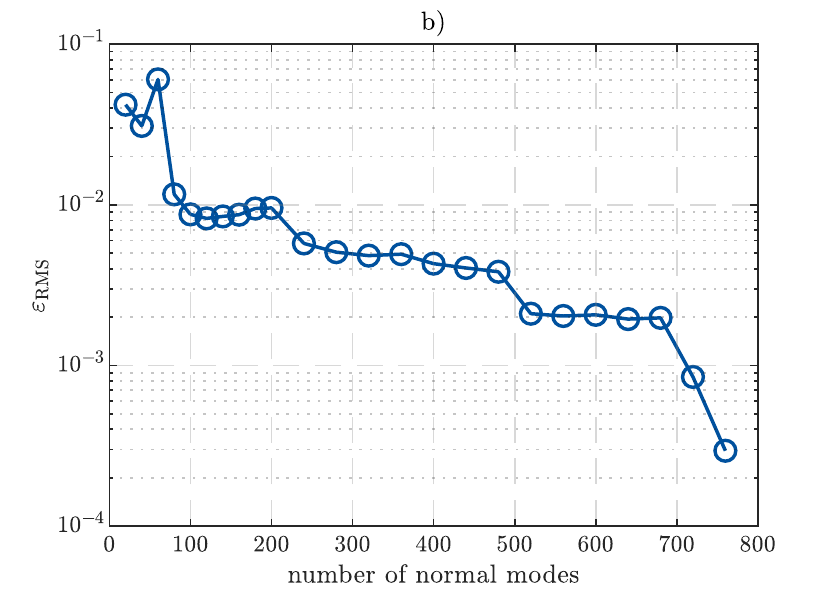}
    \end{subfigure}
    \caption{
    \MOD{Convergence with respect to number of normal modes retained in MacNeal basis: (a) representative time interval of the upper beam's tip velocity; (b) relative root-mean-square error; the representative time interval corresponds to the first five periods illustrated in \fref{velsignal_high}-top-left}
    }
    \label{fig:convergence_modal}
\end{figure}

\section{Conclusions\label{sec:conclusions}}
Overall, we conclude that the agreement between predictions and experimental results is very good.
This assessment should be viewed in the light of the challenging aspects of the problem setting, namely the non-periodic character of the response, the severe frictional impact interactions, and the relevance of the dynamic flexibility of both contacting bodies.
We attribute the remaining deviations to the uncertainty associated with hammering wear occurring during the individual stepped sine tests, and the slightly nonlinear behavior (in particular: damping) of the clamping.
We conclude that the proposed modeling and simulation approach is valid for problem settings similar to that analyzed in the present study.
Its predictive capability is limited by two aspects that were required to obtain reasonable agreement:
First, the approach relies on updating modal frequencies and damping ratios (as almost any state-of-the-art approach).
Second, we updated the actual clearance due to substantial material loss in the contact region (which may not be relevant in many vibro-impact problems).
It should be emphasized, however, that the model updating is strictly limited to measurements obtained in the (quasi-linear) regime without impacts.
In particular, we did not calibrate empirical parameters, such as a coefficient of restitution, to the measured vibration response.
At the same time the robustness and efficiency of the simulation are found to be quite acceptable.
Consequently, we are convinced that the proposed approach is a major step forward in making vibro-impact processes predictable and will be useful for the predictive design of a multitude of engineering applications.

\section*{Acknowledgments}
This work was jointly funded by the Federal Ministry of Economics and Technology (BMWi) and MTU Aero Engines AG, Germany (FKZ 03EE5041G).
The authors are grateful to MTU Aero Engines AG, Germany, for giving permission to publish this work.

\end{document}